\documentclass[reprint,aps,showpacs,twocolumn]{revtex4}
\usepackage{amsmath}
\usepackage{graphicx}
\usepackage{tikz}
\begin{document}
\newcommand{\jb}{\textbf{j} }
\newcommand{\rb}{\textbf{r} }
\newcommand{\db}{\textbf{d} }
\newcommand{\nb}{\textbf{n} }
\newcommand{\pb}{\textbf{p} }
\newcommand{\vb}{\textbf{v} }
\newcommand{\Ab}{\textbf{A} }
\newcommand{\Bb}{\textbf{B} }
\newcommand{\Eb}{\textbf{E} }
\newcommand{\Pb}{\textbf{P} }
\newcommand{\Fb}{\textbf{F} }
\newcommand{\VEC}[3]{\ensuremath{\begin{pmatrix} #1\\ #2\\ #3\end{pmatrix}}}
\newcommand{\multiX}{\mathbf{X}}
\newcommand{\multiV}{\mathbf{V}}
\newcommand{\multiPi}{\mathbf{\Pi}}

\title{2D structures of electron bunches in relativistic plasma cavities}
\author{Lars Reichwein}
\email{lars.reichwein@hhu.de}
\author{Johannes Thomas}
\author{Alexander Pukhov}
\affiliation{Institut f\"{u}r Theoretische Physik I, Heinrich-Heine-Universit\"{a}t D\"{u}sseldorf, D-40225 D\"usseldorf, Germany}

\date{\today}
\begin{abstract}
	The spatial structure of an ultra-low emittance electron bunch in a plasma wakefield blowout regime is studied.  The full Li\'{e}nard-Wiechert potentials are considered for mutual inter-particle interactions in the framework of the equilibrium slice model (ESM). This model uses the quasi-static theory which allows to solve the Li\'{e}nard-Wiechert potentials without knowledge of the electrons' history. The equilibrium structure we find is similar to already observed hexagonal lattices but shows topological defects. Scaling laws for interparticle distances are obtained from numerical simulations and analytical estimations.
\end{abstract}

\maketitle
\section{Introduction}
Plasma based electron acceleration methods are known for their high efficiency which allows to accelerate electrons up to some GeV over much shorter distances than in conventional accelerators \cite{Kostyukov2015}. In the field of laser-driven plasma  acceleration, the  wakefield is excited by an intense laser pulse with wavelength $\lambda_L$, duration $\tau$ and focal spot size $R$ \cite{Tajima1979, Esarey2009}. In plasmas with homogeneous density, the wakefield breaks as soon as the laser pulse intensity reaches a certain threshold value and the normalized laser amplitude $a_0 > 1$. If $a_0 > 4$, $R > 2\lambda_L$ and if the laser pulse perfectly fits into the first half of the plasma period, a solitary electronic cavity, called the bubble, is formed \cite{Pukhov2002, Jansen2014, Pukhov2006, Lu2007}. It is a nearly spherical region with uniform accelerating fields that propagates with almost speed of light $c$ \cite{Kostyukov2004} and traps background electrons at its tail. The major features that characterize the bubble regime are the quasi-monoenergetic spectrum of the fast electrons and a quasi-static laser pulse, which propagates many Rayleigh lengths in homogeneous plasma without significant diffraction.
Another method to excite a plasma wakefield is to use a charged particle beam with length $\sigma_z$, radius $\sigma_r$ and density $n_b$ \cite{Chen1985, Rosenzweig1988}. If the particle beam is thin $\sigma_z\approx\sqrt{2}k_p^{-1}\gg \sigma_r$ and if its density much larger than the electron plasma density, a structure similar to the bubble, the so called blow-out, is created.
In both cases, a nearly harmonic wakefield potential accelerates trapped electrons to high energies and focuses them to the axis where they form a dense electron beam - the so called beam load. 

Two promising methods to control the beam load formation are the density down-ramp and the ionization injection technique. Both methods produce witness electron beams with sub-fs temporal duration, a very high peak current of several kA, energy spreads well below 1\% and an excellent transverse emittance \cite{Huang2017, Baxevanis2017, Wang2018, Gonsalves2017, Martinez2017, Tooley2017}. The density down-ramp injection is reached by a longitudinal modulation of the plasma density with potentially extremely large gradients (also known as shock-fronts) \cite{Gonsalves2011, Swanson2017, Xu2017, Xu2017a}. The ionization injection requires a small amount of higher-Z gas, added to the gas used for acceleration \cite{Pak2010, Tochitsky2016}.  If the wakefield is driven by a short electron beam, the Trojan horse regime (THWFA) of underdense photocathode PWFA is reached \cite{Hidding2012, Hidding2012a}. It is best suited to decouple the electron bunch generation process from the excitation of the accelerating plasma cavity.  The combination of the non-relativistic intensities required for tunnel ionization ($10^{14}$ W/cm$^2$), a localized release volume as small as the laser focus, the greatly minimized transverse momenta, and the rapid acceleration leads to dense phase space packets. In homogeneous plasma they can have ultra-low normalized transverse emittance in the bulk of $\mu$m\,mrad and a minimal energy spread in the 0.1\% range \cite{Hidding2012, Chen2014}.

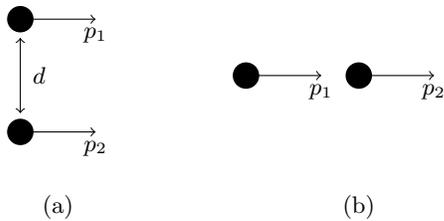
\begin{figure}[t]
		\begin{tikzpicture}
	\draw (0,0.5);
	\fill[black] (0,0) circle (5pt);
	\fill[black] (0,-1.5) circle (5pt);
	\draw[->] (0,0) -- (1,0) node[anchor = north] {$p_1$};
	\draw[->] (0,-1.5) -- (1,-1.5) node[anchor = north] {$p_2$};
	\draw[<->] (0,-0.25) -- (0,-1.25);
	\draw (0.25,-0.75) node {$d$};
	\draw (0.5,-2.5) node {(a)};

	\fill[black] (3,-0.75) circle (5pt);
	\fill[black] (4.5,-0.75) circle (5pt);
	\draw[->] (3,-0.75) -- (4,-0.75) node[anchor = north] {$p_1$};
	\draw[->] (4.5,-0.75) -- (5.5,-0.75) node[anchor = north] {$p_2$};
	\draw (4.5,-2.5) node {(b)};
	\end{tikzpicture}
	\caption{\label{fig:scheme} Schematic depiction of (a) two alongside propagating particles and (b) two particles propagating one behind the other in the same direction.}
\end{figure}

Besides present days aim to produce electron bunches with highest energies and smallest transverse  emittances, it is also important to discuss the spatial beam load structure. In some recent experiments the length, the diameter and the emittance of the beam load were measured to determine the beam quality \cite{Schnell2012, Saevert2015}. If, however, the relativistic emittance falls below a certain threshold value and if the electron energy is sufficiently low, the inter-particle interaction becomes important and starts compensating the focusing force of the wakefield.  We know that this repelling force between two alongside propagating electrons (Fig. \ref{fig:scheme} (a)) scales inversely proportional to their energy $E=\gamma m_ec^2$, where $\gamma$ is the Lorentz factor and $m_e$ is the electron mass. If both particles propagate in one behind the other (Fig. \ref{fig:scheme} (b)) the interaction force scales like $1/E^2$.  For this reason it is convenient to neglect any inter-particle interaction between accelerated electrons in the bubble regime as long as the beam load energy is in the GeV regime. For much lower energies between some tens MeV and some hundred MeV and a transverse emittance of $10^{-9}$\,m\,rad, alongside propagating neighboring electrons will repel each other such that a spatial structure, which is known as the equilibrium structure of the beam load, can be considered. To analyze this structure a suitable description of the mutual electron interaction the bubble is necessary. 

One method to describe a retarded interaction of electrons within the bubble is to superpose the wakefield known from a quasi-static model with the interaction field described by the Li\'{e}nard-Wiechert fields for relativistically moving point-like charges. However, this approach would require knowledge of the history of all electrons within the beam load. To circumvent this disadvantage it is necessary to find an explicit expression of the retarded time in terms of the actual system time, the space variables and the momentum variables. Once such an expression is found and substituted into the Li\'{e}nard-Wiechert fields, a new quasi-static interaction model is derived. In this way the approximation of the retarded time effects the predicted equilibrium structure.

One important example for an interaction bubble model incorporating an explicit expression of the retarded time is introduced in \cite{Thomas2017} where 2D and 3D equilibrium distributions are calculated on the basis of a Taylor expansion of the Li\'{e}nard-Wiechert potentials in terms of v/c. Here, the series is cut after the second order because higher orders would require terms including the electron acceleration and thus radiation effects, too. Another argument given in \cite{Thomas2017} to cut the Taylor series after the second order is that retardation effects during the interaction of charged particles become small either if the particles move much slower than the speed of light, or if the distance traveled by light in the time gap between the retarded time and actual time is large against the mean electron distance. In the scope of this approach it could be shown that the 2D equilibrium structure is similar to Wigner crystals observed in other areas of plasma physics, like e.g. dusty plasmas, while the 3D equilibrium distribution shows a completely new spatial symmetry. 

In our current work we discuss a new approach which allows us to calculate the retarded time with arbitrarily high precision and without knowledge of the electrons' history. We analyze the mutual electron interaction with a moderate energy up to some hundred MeV. At this energy level the interaction force between two alongside propagating electrons is more than a hundred times stronger than it is for electrons traveling one behind the other. Thus we subdivide the beam load into multiple slices similar to the approach in \cite{Schulte2017} and discuss the results of this equilibrium slice model (ESM) for zero transverse emittance beam loads and full Li\'{e}nard-Wiechert potentials. Afterward we compare our results to the 2D structure presented in \cite{Thomas2017} and discuss topological defects in the symmetry of the equilibrium distributions. We find that, while having a different size and more topological defects, similar hexagonal lattices as before are observed. The differences in size and number of defects can be explained by the more precise modeling of the system without Taylor expansions. An additional comparison of our numerical simulations to analytical scaling laws derived from a two particle system of alongside propagating relativistic electrons shows that the analytic scalings hold even for system with a much higher number of particles.

\section{The mathematical model}
\begin{figure}[t]
	\includegraphics[scale=.55]{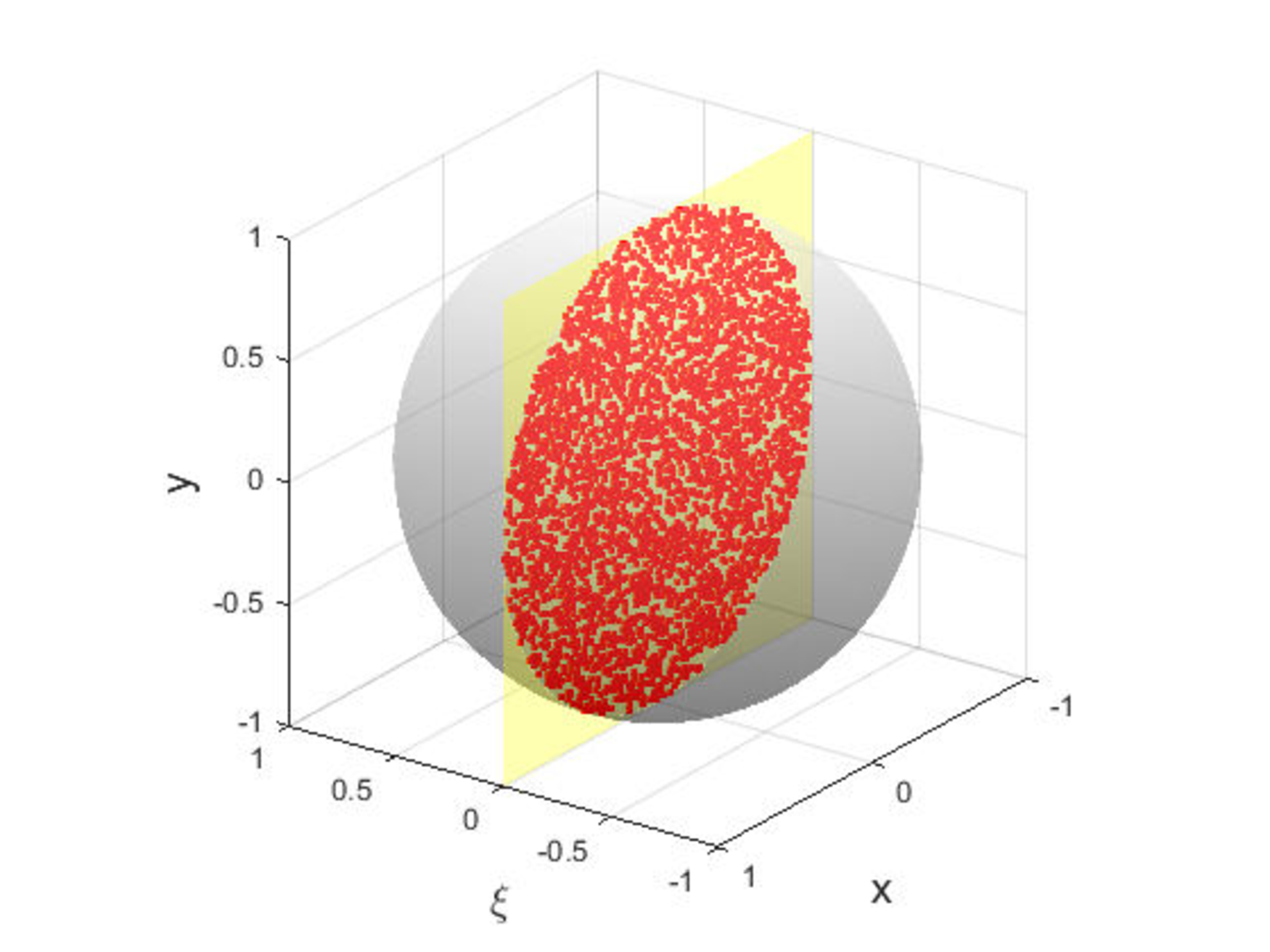}
	\caption{\label{fig:distribution}Schematic depiction of the random distribution on a circular disk inside the bubble.}
\end{figure}

In the following we derive the Hamiltonian for a system of interacting alongside accelerating relativistic electrons in external potentials in a moving coordinate system. In the scope of this model  we analyze the equilibrium structure of electrons which are distributed on a circular disk inside a 3D bubble such that $\xi=z -V_0t$ is the same for all electrons (Fig. \ref{fig:distribution}, red dots on the yellow hyper-plane). Here, $z$ is the propagation direction of the bubble, $V_0$ is the bubble velocity and $\xi$ is the particles' longitudinal position inside the bubble in the moving system. The equilibrium structure is found by numerical simulations minimizing the Hamiltonian for the special case that the external potentials are known from the strongly simplified quasi-static 3D bubble model for electron acceleration in homogeneous plasma \cite{Kostyukov2004}. In this model the acceleration in the direction of propagation is just due to the external electric field $E_z=\partial\Psi/\partial\xi$, where $\Psi=\varphi-A_z=(x_i^2+y_i^2+\xi_i^2)/8$ is the wakefield potential.

The basic mathematical model for the electron-electron interaction are the retarded Li\'{e}nard Wiechert potentials $\varphi_{LW}$ and  $\Ab_{LW}$ which originate from point-like particles $i$ with charge $q_i$, position $\rb_i$ and velocity $\vb_i$ at retarded time $t_i$ and are measured at time $t$ and position $\rb$
\begin{eqnarray}
\varphi_{LW}=\sum_{i=1}^n\frac{\Lambda q_i}{|\rb(t)-\rb_i(t_i)|-\vb_i(t_i)\cdot [\rb(t)-\rb_i(t_i)]},\\
\Ab_{LW}=\sum_{i=1}^n\frac{\Lambda q_i\vb_i(t_i)}{|\rb(t)-\rb_i(t_i)|-\vb_i(t_i)\cdot [\rb(t)-\rb_i(t_i)]}.
\end{eqnarray}
In this form we have normalized time to the inverse plasma frequency $\omega_p^{-1}=\sqrt{4\pi e^2 n_e/m_e}^{-1}$, lengths to the inverse plasma wave number $k_p^{-1}=c/\omega_p$, kinetic momenta to $m_e c$, energy to $m_ec^2$, fields to $m_ec\omega_p/e$, charges to the elementary charge $e$, masses to the electron rest mass $m_e$ and potentials to $m_ec^2/e$. The pre-factor $\Lambda = r_e/\lambda_{pe}$ consists of the classical electron radius $r_e=2\pi e^2/(m_e c^2)$ and the plasma wavelength $\lambda_{pe}$.

To model the inter-particle interaction by the Li\'{e}nard-Wiechert potentials it is necessary to know the history of all electrons within a slice or to find an explicit expression of the retarded time $t_i$ in terms of the actual system time, the space variables and the momentum variables. In the following we discuss the idea that all electrons in the slice are accelerated coherently such that they have the same (time-dependent) $p_z$ and $\xi$ but different constant radial positions $r_i$ and thus $p_r=0$. This approach is equivalent to the assumption that the electron ensemble is already in equilibrium and that we seek to find its spatial structure by minimizing the total energy of the system.

Since we assume that the kinetic energy of our electrons is much larger than their rest energy, $p_z\ll \gamma_0$ and the solutions of the equations of motion of a single not-interacting test electron can be expressed in terms of the smallness parameter $\epsilon_0 = 1/(2\gamma_0^2)$. Written up to the second order in $\epsilon_0$ the solutions are
\begin{equation}
p_z = p_0 - \frac{\xi_0}{2}(t-t_0) - \frac{\epsilon_0}{4}(t-t_0)^2,~z = z_0 + \int_{t_0}^t\frac{p_z}{\gamma}dt'.
\end{equation}
For boundary conditions $t_0=0$, $z(t_0)=\xi(t_0)=\xi_0$ and  $p_z(t_0)=p_0$ our approximations result in implicit expressions for the positions and velocities of our electrons:
\begin{eqnarray}
\rb_i(t)=\begin{pmatrix}
x_{i,0} \\ 
y_{i,0} \\ 
\xi_0 + \int_0^t v_{z}dt'
\end{pmatrix}, ~&&
\vb(t)=\frac{p_{z}}{\gamma}\hat{e}_z.
\end{eqnarray}
In general, the retardation of time is given by $t_{\text{ret}}=t-|\rb_i(t)-\rb_j(t_{\text{ret}})|$, but for $t=t_0=0$ the retarded time of the $j$-th particle simplifies to
\begin{eqnarray}
t_j&=&-|\rb_i(t)-\rb_j(t_j)|\notag\\
&=&-\sqrt{\Delta x_{ij}^2+\Delta y_{ij}^2+\left(\int_0^{t_j}v_{z}dt\right)^2},
\end{eqnarray}
where $\Delta x_{ij}=x_{i,0}-x_{j,0}$ and $\Delta y_{ij}=y_{i,0}-y_{j,0}$ describe the time-independent distance of particle $i$ at time $t$ to particle $j$ at time $t_j$ in $x$- or $y$-direction, respectively. 

Within this approach the particle positions are known analytically but the retarded times $t_j$ still need to be computed numerically. The retarded Li\'{e}nard Wiechert potentials, created by the $j$-th particle and seen by the $i$-th particle at time $t$, simplify to
\begin{eqnarray}
\varphi_{ij}&=&\frac{\Lambda q_j\gamma_{j}}{-\gamma_{j}t_{j}+p_{zj}\int_0^{t_{j}}v_{z}dt},\\
\Ab_{ij}&=&\frac{\Lambda q_j p_{zj}}{-\gamma_{j}t_{j}+p_{zj}\int_0^{t_{j}}v_{z}dt}\hat{e}_z.
\end{eqnarray}
From now on we will use the index $j$ to indicate that we are using a variable that is given at the retarded time $t_j$ and  index $i$ when dealing with the laboratory time $t=0$. Then, the corresponding $n$-particle system Lagrangian is
\begin{eqnarray}
	L&\approx&\sum_{i=1}^n\left[-\frac{1}{\gamma_i}+q_i\vb_i\cdot\Ab(\rb_i)-q_i\varphi(\rb_i)\right]\notag
	\\&-&\sum_{i>j}\left(1-\frac{p_{iz}p_{jz}}{\gamma_i\gamma_j}\right)q_i\varphi_{ij}
\end{eqnarray}
because all electrons move in one direction and have the same momentum. The first term in brackets is the Lagrangian of a free particle, while the second and third term describe the coupling of the $i$-th electron to the external potential. The last sum incorporates the retarded electron-electron interaction.  

To calculate the Hamiltonian we need to know the canonical momentum of each particle $\pi_i$, which in principle is the strict derivative of the Lagrangian with respect to the velocity $\vb_i$. Instead we will, analogous to the calculation by Landau-Lifshitz \cite{Landau2009}, consider the term describing the interaction as perturbation. Then
\begin{eqnarray}
	\pi_i=\frac{\partial L}{\partial \vb_i}\equiv \pb_i+q_i\Ab(\rb_i),~&& \vb_i=\frac{\pi_i-q_i\Ab(\rb_i)}{\gamma_i}
\end{eqnarray}
and we only have to change the signs of the interaction parts. Furthermore, we perform a canonical transformation to the co-moving frame, described by the coordinate $\xi$. Then
\begin{eqnarray}
H&=&\sum_{i=1}^n [\gamma_i +q_i\Psi(\rb_i) -p_{iz}\notag
\\ &+&\sum_{i>j}\frac{r_e}{\lambda_{pe}}\left(1-\frac{p_{iz}p_{jz}}{\gamma_i\gamma_j}\right)q_i\varphi_{ij}]. \label{HAM}
\end{eqnarray}
Since we will be using an iterative algorithm to find the energetic minimum of the system, we need the gradient of our Hamiltonian, which is given by
\begin{eqnarray}
\nabla_{i\bot} H = \frac{1}{2}\binom{x_{i0}}{y_{i0}}+\frac{r_e}{\lambda_{pe}}\left(\sum_{j\neq i}\left(1-\frac{p_{iz}p_{jz}}{\gamma_i\gamma_j}\right)\nabla_{i\perp}\varphi_{ij}\quad\right. \label{DHAM}\\ 
\quad	-\left.\sum_{j\neq i}\varphi_{ij}\left(\frac{p_{iz}}{\gamma_i}\frac{\partial}{\partial p_{jz}}\frac{p_{jz}}{\gamma_j}\frac{\partial p_{jz}}{\partial t_j}\nabla_{i\perp}t_{j}+\frac{p_j}{\gamma_j}\nabla_{i\perp}\frac{p_i}{\gamma_i}\right)\right),\notag
\end{eqnarray}
where $\nabla_{i\bot}=\hat{e}_x\partial_{x_i} +\hat{e}_y\partial_{y_i}$.
We will cover the numerical procedure in further detail in the section \ref{NUM}, where we will also present our numerical findings. In the next section we will calculate the dependencies of the inter-particle distance in equilibrium on our simulation parameters analytically, such that we can compare these to our simulation results.

\section{Scaling Laws}
In the following we derive scaling laws for the mean inter-particle distance $\Delta r$ depending on the particles' energy and the plasma wavelength. To find an analytic expression, we consider the interaction of two equally charged  alongside propagating relativistic particles which experience two counteracting  forces along their separation direction. The first force is the repelling interaction force $\Fb_\bot$ which can be calculated from the Li\'{e}nard-Wiechert potentials. The second force is an external force $\Fb_\text{ext}$ which focused the particles to the origin but does not accelerate them in direction of motion. Similar to the mathematical model we assume that the particles are resting in their equilibrium positions and calculate their distance.

Starting from the scalar potential $\varphi$, we have
\begin{eqnarray}
\varphi &=& \Lambda\frac{q_1 q_2}{|\rb_1(t)-\rb_2(t_{2})|-\vb_2(t_{2})\cdot (\rb_1-\rb_2(t_{2}))}\notag\\
	&=&\Lambda\frac{q_1 q_2}{d_{12}}\cdot\frac{1}{1-\nb_{2}\cdot\vb_2(t_{2})},
\end{eqnarray}
where
\begin{equation}
	\nb_{2}=\frac{\db_{12}}{d_{12}}, \quad t_{2}=-d_{12}, \quad d_{12}=|\rb_1(t)-\rb_2(t_{2})|.
\end{equation}
Further, we have
\begin{eqnarray}
	&\Fb_1 = -q_1\nabla\varphi +q_1\vb_1\times(\nabla\times\Ab),\notag \\
	&\vb_1 = v_1 \hat{\textbf{e}}_z = v_2\hat{\textbf{e}}_z = \vb_2 =\text{const.}
\end{eqnarray}
so that
\begin{eqnarray}
&\Fb_\perp = \Fb_1 = -\Fb_2 = -q_1\left( 1-\vb_1^2\right)\VEC{\partial_x}{\partial_y}{0}\varphi_{12}, \\
&\rb_2(t_{2}) = \rb_2(t) -d_{12}\vb_1. \label{r2}
\end{eqnarray}
It is $\rb_1(t)-\rb_2(t)\perp\vb_1$ so that the retarded time becomes
\begin{eqnarray}
t_{2}&=& -\vert\rb_1(t)-\rb_2(t)\vert\cdot\gamma = -d\gamma. \label{t2}
\end{eqnarray}
We consider electrons with velocity $v_z\approx 1$, thus $v=|\vb_1|=|\vb_2|\approx 1$ and $q_1=q_2=-1$. Further, we assume that the electrons are located on the $x$-axis. Then, with Eq.(\ref{t2}) and Eq.(\ref{r2}) the interaction force is
\begin{eqnarray}
	F_x&\approx& -\frac{\Lambda}{\gamma^2}\frac{\partial}{\partial x}\frac{1}{d_{12}-(z_1-z_2(t_{2}))v}.
\end{eqnarray}
where
\begin{equation}
	d_{12}=|\rb_1-\rb_2|\gamma=d\gamma, \quad ~d=x_1-x_2.
\end{equation}
Since $z_1(t) = z_2(t)$, Eq.(\ref{r2}) gives
\begin{equation}
z_1 -z_2(t_{2}) = d\gamma v, \quad d_{12} -(z_1-z_2(t_{2}))v = \frac{d}{\gamma}
\end{equation}
and thus 
\begin{equation}
 F_x=-\frac{\Lambda}{\gamma^2}\frac{\partial}{\partial x}\frac{\gamma}{d}= \frac{\Lambda}{\gamma}\frac{1}{d^2}.
\end{equation}
In equilibrium the magnitude of this force must be equal to the magnitude of the external force. In the bubble regime $|\Fb_\text{ext}|=r/2$, where $\Delta r = d= 2r$ due to the symmetry of the system such that
\begin{equation}
	\Delta r = \sqrt[3]{\frac{4 r_e}{\lambda_{pe} \gamma}}.
\end{equation}
in normalized units and 
\begin{equation}
	\Delta r  = \sqrt[3]{\frac{r_e }{2\pi^3}}\left(\frac{\lambda_{pe}}{\sqrt{\gamma}}\right)^{2/3} \propto p^{-1/3}\lambda_{pe}^{2/3} \label{scaling}
\end{equation}
in cgs units. This scaling holds for all systems of two alongside propagating electrons with constant velocity and constant distance. In the mathematical model of the two-dimensional beam load slice these conditions are fulfilled in parts because we assumed that all electrons have fixed radial positions and move with the same longitudinal velocity. This velocity, however, is not constant because we consider all particles to be accelerated coherently in the bubble in longitudinal direction. Further, the physical system of the slice is an $n$-particle system but the interaction is modeled as a sum of two-particle interactions. Thus we expect that the scaling (\ref{scaling}) holds for an arbitrary high but fixed number of electrons. The dependency on $n$ in turn must be determined numerically and will so in the following section.
\\

Another important point regarding scaling laws is a comparison of our new approach to the one presented in \cite{Thomas2017}.
Here, a Taylor expansion of the Li\'{e}nard-Wiechert potentials in terms of $v/c$ was used to find the radially repelling interaction force
\begin{equation}
	F_{TR} \propto \Lambda\frac{1}{\Delta r^2}\left(\frac{1}{\gamma}+\frac{v^2}{2}\right),
\end{equation}
while the external force $F_{\text{ext}}=r/2$ is the same as in our model. In the limit $v^2\approx 1$ the equilibrium condition is
\begin{equation}
	\Delta r_{TR}^3 \propto 2\Lambda\left(\frac{1}{\gamma}+\frac{1}{2}\right)
\end{equation}
so that 
\begin{equation}
	\Delta r_{TR}\propto \sqrt[3]{\frac{1}{\gamma}+\frac{1}{2}}\cdot \lambda_{pe}^{2/3} \label{scalTR}
\end{equation}
in cgs units. If we consider large electron energies $\gamma \gg 2$, the square root in Eq.(\ref{scalTR}) can be expanded in terms of $\gamma^{-1}$ and the scaling of the mean electron distance in a 2D slice in the beam load becomes
\begin{equation}
	\Delta r_{TR}\propto 2^{-2/3}\lambda_{pe}^{2/3} +\frac{2^{2/3}}{3} \lambda_{pe}^{2/3}\gamma^{-1}.
\end{equation}
If we compare $\Delta r_{TR}$ to the scaling law in Eq.(\ref{scaling}) we see that
\begin{eqnarray}
\frac{\Delta r_{TR}}{\Delta r_{LW}} \propto \sqrt[3]{\frac{\gamma}{2}} +\frac{1}{3}\left(\frac{2}{\gamma}\right)^{2/3}.
\end{eqnarray}
For high electron energies the second term approaches zero and
\begin{equation}
	\frac{\Delta r_{TR}}{\Delta r_{LW}} \propto \sqrt[3]{\gamma}.
\end{equation}
This estimation clearly shows that the mean electron-equilibrium distances in the interaction model \cite{Thomas2017} are more than one order of magnitude larger than those predicted by our theory if the electrons have an energy in the near GeV regime. For energies in the lower MeV regime the difference between the distances predicted by the models is rather small and definitely in the same order.

\section{The 2D equilibrium state}\label{NUM}
In this section we present the numerical method we use to minimize the Hamiltonian in Eq.(\ref{HAM}) and discuss the equilibrium structure in the 2D beam load slices we find. In this context we compare the predictions of the scaling laws from the previous section to our numerical simulations for a fixed number of electrons and discuss topological defects. Furthermore, the scaling of the mean particle distance in the equilibrium structure with the number of electrons is analyzed.
\\

For our simulations we choose the steepest descent method to find the equilibrium structures. It is an iterative algorithm
\begin{equation}
	\multiX^{k+1}=\multiX^k-\nabla_k H\cdot \Delta t, 
\end{equation}
that shifts the particles' positions $\multiX^k=(\rb_1^k,\dots,\rb_n^k)$ at time step $k$ contrary to the direction of the gradient known from Eq.(\ref{DHAM}). The step size $\Delta t$ is an appropriately chosen in order to obtain the distribution $\multiX_0=(x_{1,0},y_{1,0},\dots,x_{n,0},y_{n,0})$, such that $(\nabla_\multiX H)[\multiX_0]$ vanishes.

\begin{figure}
\includegraphics[scale=.3]{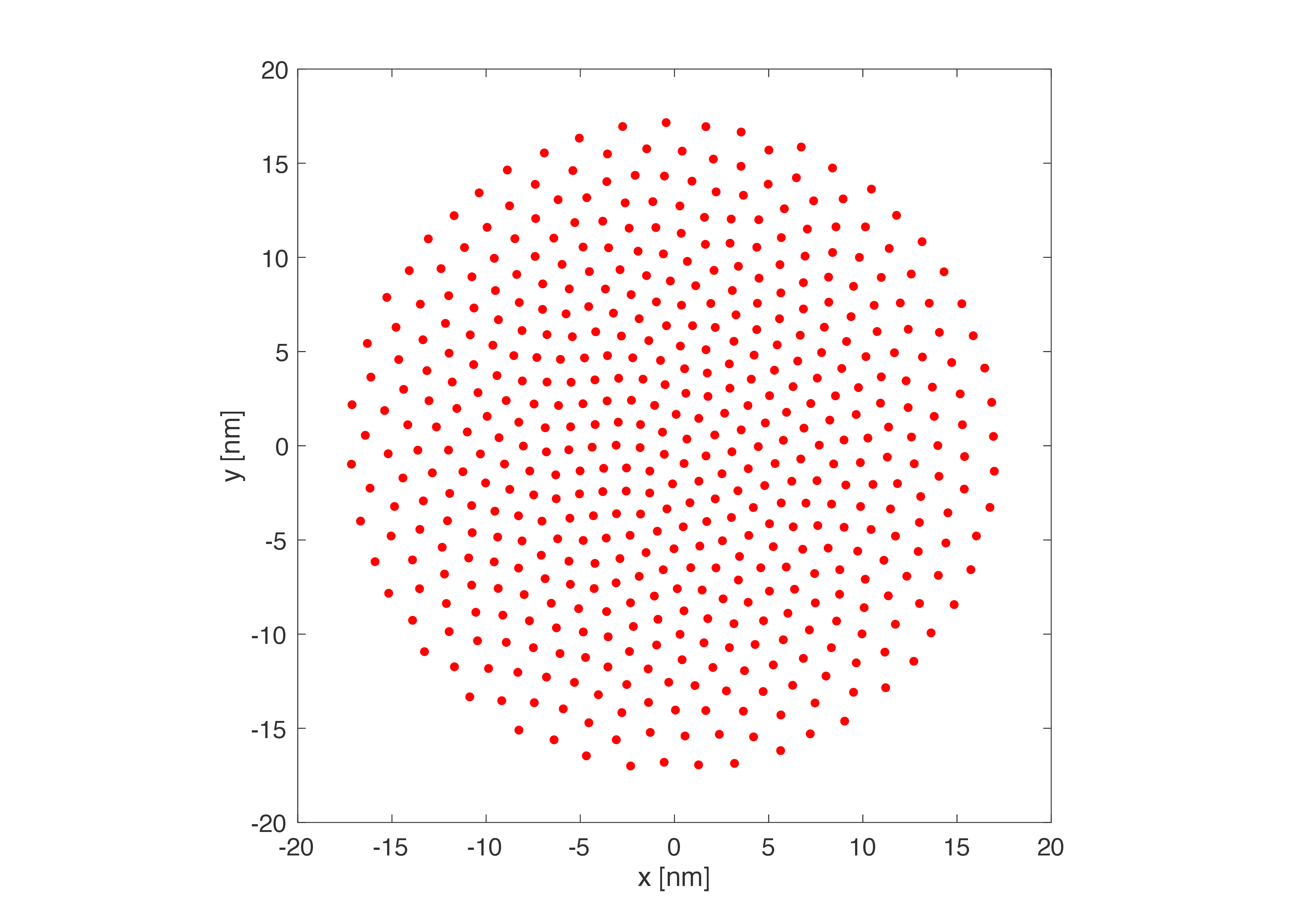}
\caption{\label{fig:final}Final distribution of a simulation with a momentum of $p=125$ MeV/c and $n=1000$ electrons with $\lambda_{pe}=0.01$ cm.}
\end{figure}

\begin{figure}
\includegraphics[scale=.3]{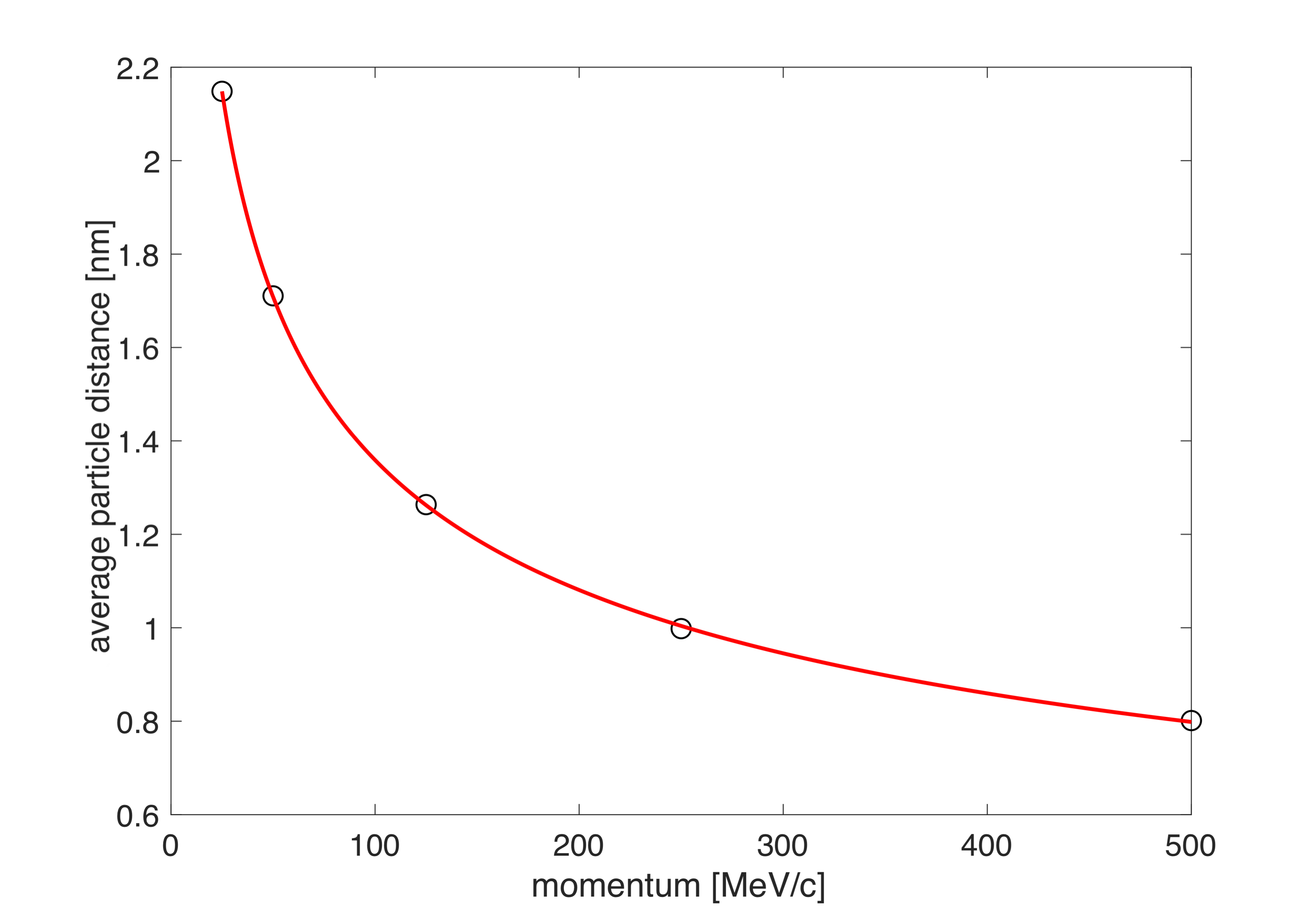}
\caption{\label{fig:momenta}Dependence of the mean particle distance for different momenta and $n = 1000$ electrons  with $\lambda_{pe}=0.01$ cm.}
\end{figure}

\begin{figure}
	\includegraphics[scale=.3]{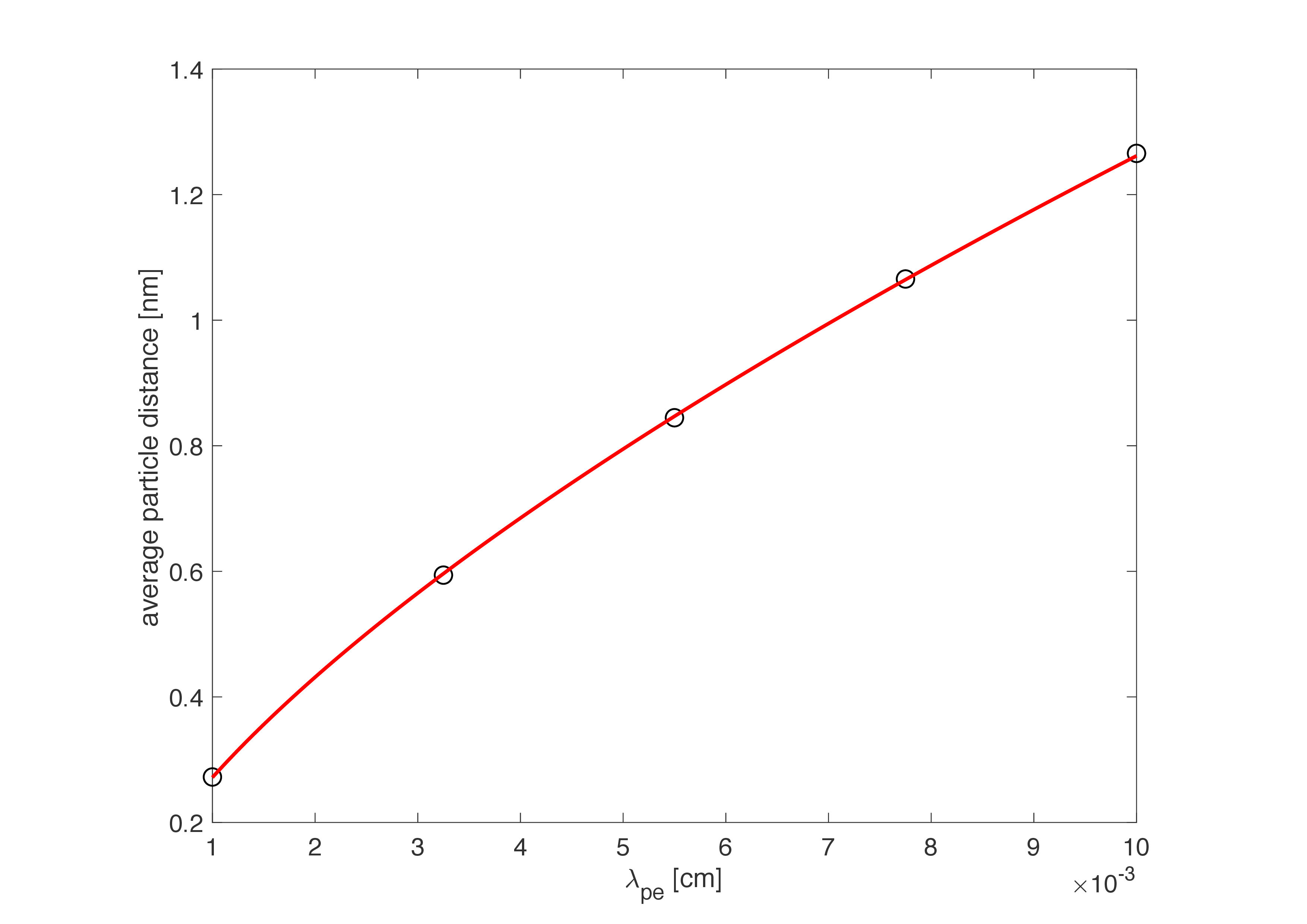}
	\caption{\label{fig:wave}Dependence of the mean particle distance for different plasma wavelengths. The simulations were done with $n = 1000$ electrons at 125 MeV/c.}
\end{figure}

We distribute a fixed number of 1000 electrons randomly on a circular disk with fixed $\xi$ coordinate (see Fig. \ref{fig:distribution}) inside the bubble and see hexagonal lattices as spatial equilibrium distribution (Fig.~\ref{fig:final}), analogous to \cite{Thomas2017}.
In our simulations we vary the electron momenta between 25 MeV/c and 500 MeV/c and observe a decrease in the mean particle distance $\Delta r$ with increasing momentum which scales like
\begin{equation}
\Delta r_{\text{phy}}\propto p^{-1/3},
\end{equation}
as can be seen from the fit in Fig. \ref{fig:momenta}. Here, $\Delta r_{\text{phy}}$ is the average distance between the nearest neighbors in the lattice in cgs units given by a Delaunay triangulation \cite{Radzvilavicius2011}. 
Regarding the scaling of $\Delta r$ in dependence of the plasma wavelength $\lambda_{pe}$, Fig. \ref{fig:wave} shows that
\begin{equation}
	\Delta r_{\text{phy}}\propto \lambda_{pe}^{2/3}.
\end{equation}
These numerical scalings are in excellent agreement with our analytical approach in Eq.(\ref{scaling}) in the section above. Furthermore, they coincide with the findings of \cite{Thomas2017}, although our new approach yields the correct prefactors.

\begin{figure}
	\includegraphics[scale=.3]{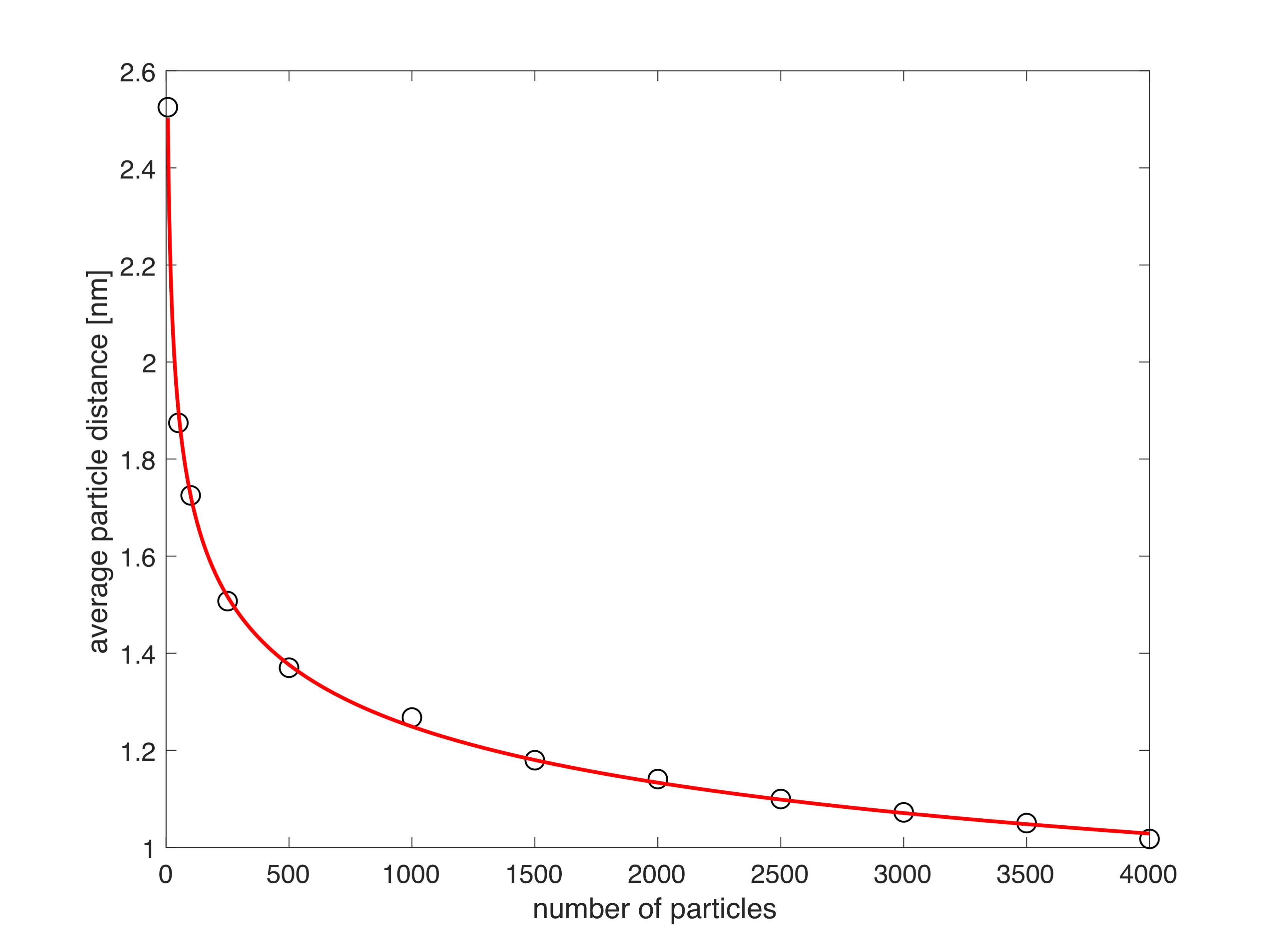}
	\caption{\label{fig:particles}Dependence of the mean particle distance for a different number of electrons and constant momentum of $125$ MeV/c with $\lambda_{pe}=0.01$ cm.}
\end{figure}

For an increasing number of electrons we observe a decreasing trend for the mean distance between particles which scales like
\begin{equation}
	\Delta r_{\text{phy}}\propto n^{-0.14}.
\end{equation}
The corresponding fit in Fig.~\ref{fig:particles} shows an excellent correspondence.
Due to the confinement of the electrons in the harmonic bubble potential and their endeavor to repel each other, a high amount of stress onto the lattice structure is produced. In order to reduce this stress, topological defects can arise. A topological defect is a deviation in the number of nearest neighbors from the reference value. We find these again with a Delaunay triangulation \cite{Radzvilavicius2011} such that we get the topological charge $Q_{top}$ of each electron:
\begin{equation}
Q_{\text{top}}=\tilde{Q}-Q_{nn}.
\end{equation}
Here $\tilde{Q}$ is the reference value (in the case of our ideally hexagonal lattice $\tilde{Q}=6$) and $Q_{nn}$  is the actual number of nearest neighbors. A topological defect exists if $Q_{\text{top}}\neq 0$.
\begin{figure}
\includegraphics[scale=.55]{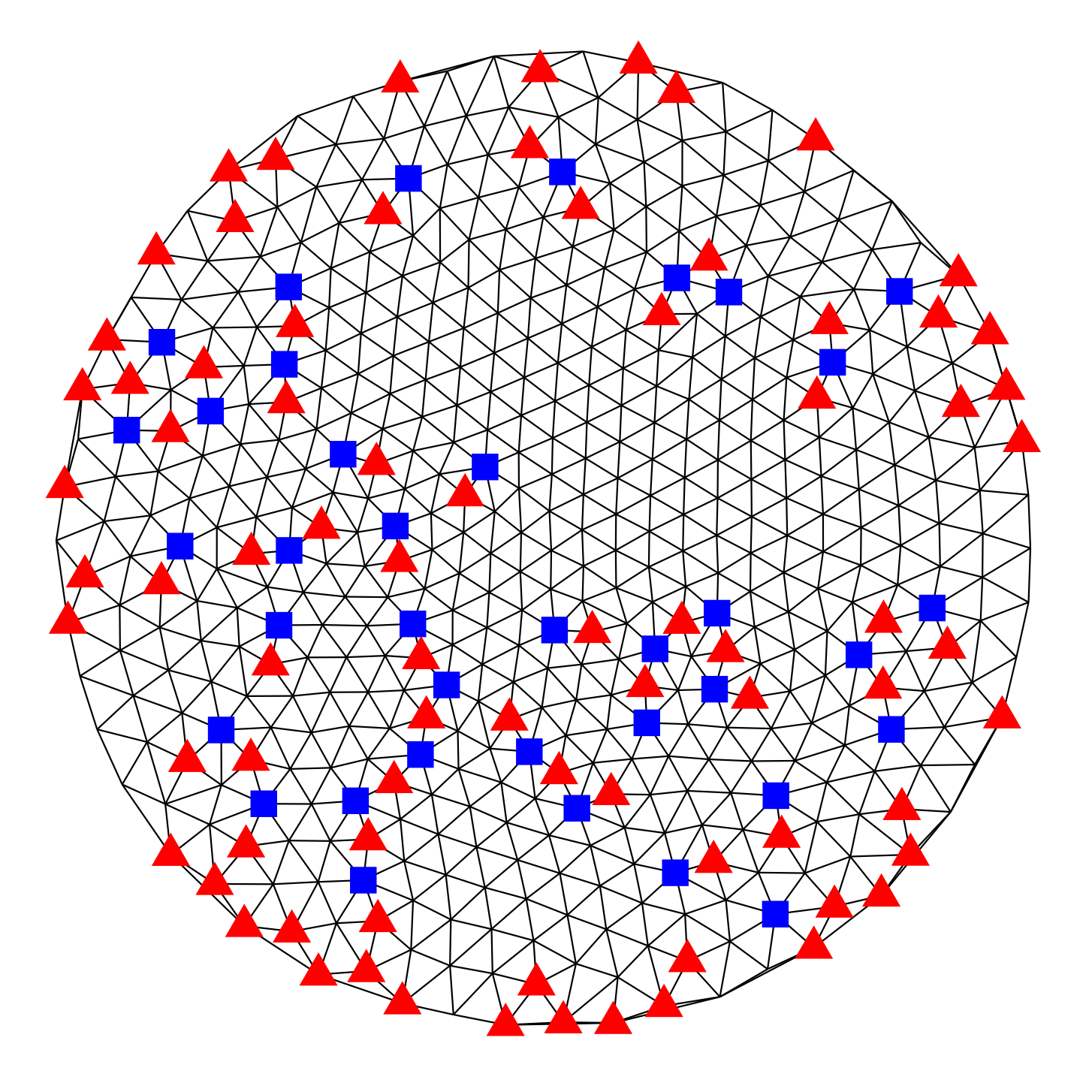}
\caption{\label{fig:defects}Topological defects in a lattice consisting of 500 electrons. Marked as red triangles are electrons with a topological charge of $-1$, marked as blue squares are particles with topological charge $+1$.}
\end{figure}
Especially eye-catching is the formation of defect chains (alternating red and blue dots in Fig.~\ref{fig:defects}): electrons with only five nearest neighbors hold at least one neighbor that holds seven nearest neighbors of its own, etc. 
The increase of topological defects with increasing momentum is i.a.~explained by the higher stress put onto the lattice.

\begin{figure}
\includegraphics[scale=.3]{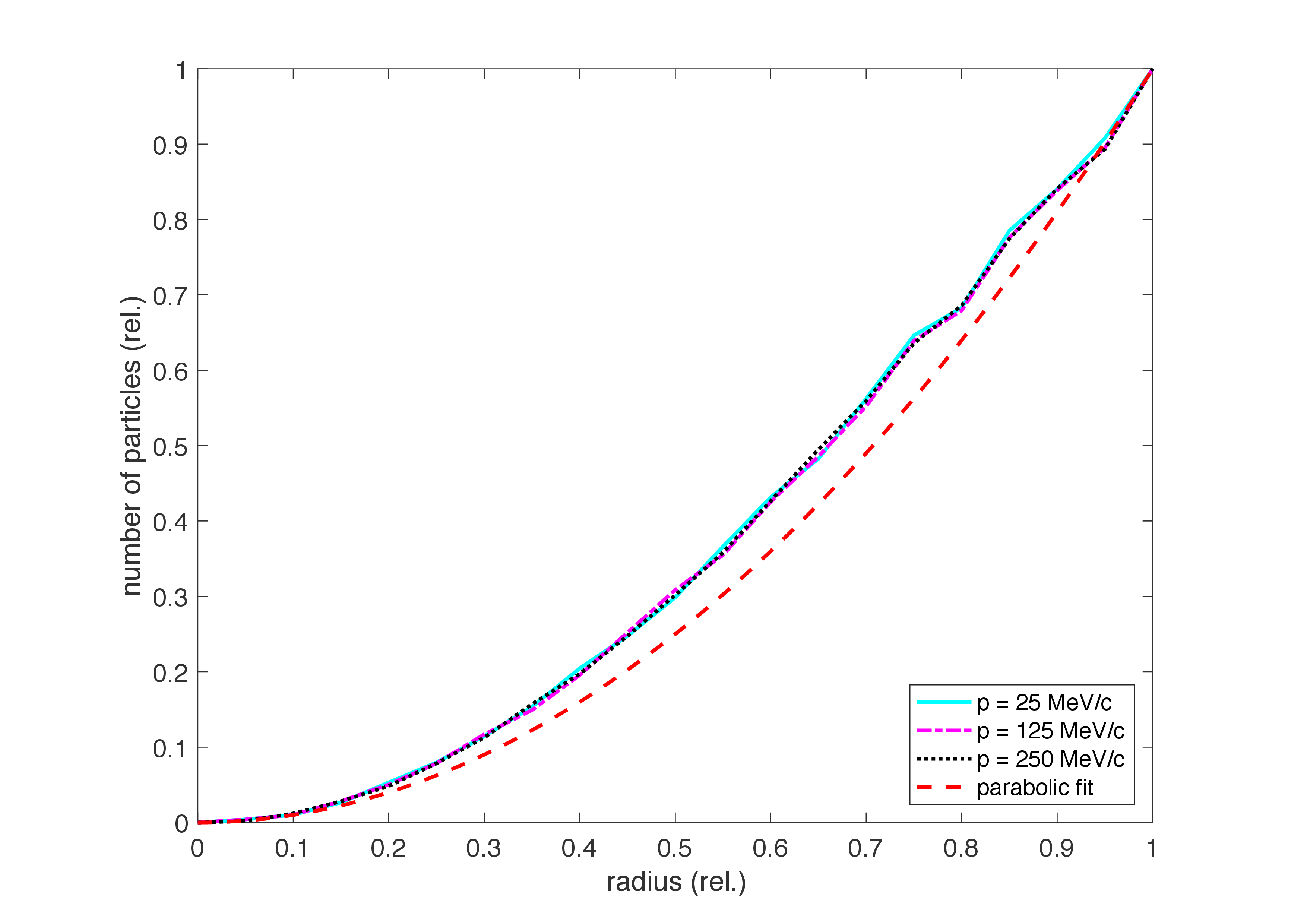}
\caption{\label{fig:momentum_multi}Depiction of the number of electrons in dependence of the observed radius of the final distribution for different momenta. The dashed line depicts the reference curve of a constant surface density.}
\end{figure}

Regarding the density of our 2D distribution we expect that the number of particles grows quadratically with the radius of the distribution for a constant density (Fig.~\ref{fig:momentum_multi}). However, we can see that there is a higher electron density in the middle of the distribution, which declines towards the edge. This is mainly due to the fact that a transition between the hexagonal structure of the lattice and the circular symmetry of the confining potential is needed. Differences between the density gradients for different momenta or number of electrons, respectively, are negligible.

Different to \cite{Thomas2017}, where mean inter-particle distances in the range of $\Delta r_{\text{phy}} \approx 1$ nm were observed for $p=270$ MeV/c and $n=4000$ electrons at a plasma wavelength of $\lambda_{pe}=11$ $\mu$m, our current approach predicts an average distance of $\Delta r_{\text{phy}} \approx 0.18$ nm. This is in good agreement with our analytic comparison in the previous section. For lower energies and larger plasma wavelengths, we observe inter-particle distances in the nanometer range  (Fig. \ref{fig:momenta}), since on this scale electric and magnetic fields do not compensate completely. Since we are able to calculate $t_j$ with higher accuracy, we incorporate more of the retardation effects and therefore observe more deviations from a perfect hexagonal lattice than in \cite{Thomas2017}, i.e. more topological defects arise.

\section{Conclusion}
We have presented a new method to find an explicit expression of the retarded time in terms of the actual system time, the space variables and the momentum variables in order to avoid needing knowledge of the history of all electrons in a 2D beam load slice.
After substituting the retarded time into the Li\'{e}nard-Wiechert potentials we introduced a new equilibrium slice model (ESM) for relativistically moving point-like test-electrons as a superposition of the Li\'{e}nard-Wiechert fields and the confining field from a quasi-static analytical bubble model. Since the model for the retarded time defines the interaction of the electrons, it also determines the equilibrium structure. 

We derived scaling laws from a two-particle system of alongside propagating relativistic electrons. These scalings fit perfectly to our numerical results even for a much higher number of particles. The equilibrium structure for many particle systems is a hexagonal lattice, similar to the ones observed \cite{Thomas2017}.
However, our new approach yields smaller mean electron-electron distances. In the context of scaling laws we also showed that the difference between the distances predicted by models scales like $\sqrt[3]{\gamma}$. This is a moderate deviation for energies up to some hundred MeV. Finally, we discussed the existence of topological defects as a mean of reducing the stress onto the lattice, which is important for higher energies since a transition between the hexagonal lattice structure and the parabolic confinement of the external field needs to be made.

\begin{acknowledgments}
This work has been supported in parts by DFG project PU-213 and BMBF project 05K2016.
\end{acknowledgments}
\bibliographystyle{prsty}
\bibliography{paper_ref}
\end{document}